\documentclass[a4paper, 11pt]{article}\usepackage[]{graphicx}\usepackage[]{xcolor}
\makeatletter
\def\maxwidth{ %
  \ifdim\Gin@nat@width>\linewidth
    \linewidth
  \else
    \Gin@nat@width
  \fi
}
\makeatother

\definecolor{fgcolor}{rgb}{0.345, 0.345, 0.345}

\usepackage{framed}
\makeatletter
 {\par\unskip\endMakeFramed%
 \at@end@of@kframe}
\makeatother

\definecolor{shadecolor}{rgb}{.97, .97, .97}
\definecolor{messagecolor}{rgb}{0, 0, 0}
\definecolor{warningcolor}{rgb}{1, 0, 1}
\definecolor{errorcolor}{rgb}{1, 0, 0}
\newenvironment{knitrout}{}{} % an empty environment to be redefined in TeX

\usepackage{alltt}
\usepackage[T1]{fontenc}
\usepackage[utf8]{inputenc}
\usepackage[english]{babel}
\usepackage{graphics}
\usepackage{amsmath, amssymb}
\usepackage{doi} % automatic doi-links
\usepackage[round]{natbib} % bibliography
\usepackage{booktabs} % nicer tables
\usepackage[title]{appendix} % better appendices
\usepackage{nameref} % reference appendices with names
\usepackage[onehalfspacing]{setspace} % more space
\usepackage[labelfont=bf,font=small]{caption} % smaller captions
\usepackage{array}

\usepackage{geometry}
\newcommand\longtitle{Normalized power priors always discount historical data}
\newcommand\shorttitle{Normalized power priors always discount} % if longtitle too long, change here
\newcommand\subtitle{}
\newcommand\longauthors{Samuel Pawel\textsuperscript{$\star$}, Frederik Aust\textsuperscript{$\dagger$}, Leonhard Held\textsuperscript{$\star$}, Eric-Jan Wagenmakers\textsuperscript{$\dagger$}}
\newcommand\shortauthors{S. Pawel, F. Aust, L. Held, E.-J. Wagenmakers} % if longauthors too long, change here
\newcommand\affiliation{
  $\star$ Department of Biostatistics, University of Zurich \\
  $\dagger$ Department of Psychological Methods, University of Amsterdam
}
\newcommand\mail{samuel.pawel@uzh.ch}
\title{
  \vspace{-2em}
  \textbf{\longtitle} \\
  \subtitle
}
\author{
  \textbf{\longauthors} \\
  \affiliation \\
  E-mail: \href{mailto:\mail}{\mail}
}
\date{June 26, 2023}

\usepackage{hyperref}
\hypersetup{
  bookmarksopen=true,
  breaklinks=true,
  pdftitle={\shorttitle},
  pdfauthor={\shortauthors},
  pdfsubject={},
  pdfkeywords={},
  colorlinks=true,
  linkcolor=blue,
  anchorcolor=black,
  citecolor=blue,
  urlcolor=black,
}

\usepackage{fancyhdr}
\newcommand{\ie}{i.\,e.,\,} % i.e.
\DeclareMathOperator{\Nor}{N} % Normal
\DeclareMathOperator{\Be}{Be} % Beta
\DeclareMathOperator{\BetaBin}{BeBin} % Beta
\newcommand{\given}{\,\vert\,} % Given
\newcommand{\that}{\hat{\theta}} % Effect estimate
\newcommand{\Xb}{\mathbf{X}}
\newcommand{\Xbt}{\mathbf{X}^\top}

\newcommand{\yb}{\mathbf{y}}

\newcommand{\thetab}{\boldsymbol{\theta}}

\newcommand{\varepsilonb}{\boldsymbol{\varepsilon}}

\IfFileExists{upquote.sty}{\usepackage{upquote}}{}
\begin{document}
\maketitle

%% Abstract
%% -----------------------------------------------------------------------------
\begin{center}
  \begin{minipage}{13cm} {\small
      \rule{\textwidth}{0.5pt} \\
      {\centering \textbf{Abstract} \\
        Power priors are used for incorporating historical data in Bayesian
        analyses by taking the likelihood of the historical data raised to the
        power $\alpha$ as the prior distribution for the model parameters. The
        power parameter $\alpha$ is typically unknown and assigned a prior
        distribution, most commonly a beta distribution. Here, we give a novel
        theoretical result on the resulting marginal posterior distribution of
        $\alpha$ in case of the normal and binomial model. Counterintuitively,
        when the current data perfectly mirror the historical data and the
        sample sizes from both data sets become arbitrarily large, the marginal
        posterior of $\alpha$ does not converge to a point mass at $\alpha = 1$
        but approaches a distribution that hardly differs from the prior. The
        result implies that a complete pooling of historical and current data is
        impossible if a power prior with beta prior for $\alpha$ is used.}
      \rule{\textwidth}{0.4pt} }
\end{minipage}
\end{center}

\section{Introduction}
Power priors are a class of prior distributions which can be used for
incorporation of historical data in Bayesian analysis of current data
\citep{Chen2000}. The basic idea is to use the likelihood of the historical data
raised to the power of $\alpha$ as the prior distribution for the model
parameters $\theta$. This leads to the posterior distribution of $\theta$
borrowing information from both the current and the historical data, typically
resulting in an information gain compared to an analysis of the current data in
isolation. The power parameter $\alpha$ is usually restricted to the interval
between zero and one, thereby determining how much the historical data are
discounted and enabling a quantitative compromise between the extreme positions
of completely trusting ($\alpha = 1$) and completely ignoring them
($\alpha = 0$).
% Since their introduction, power priors have been used in various applications
% where historical data are available, for instance, clinical trials,
% environmental science, or psychometrics \citep[see][for some case
% studies]{Ibrahim2015}.

In practice, the power parameter $\alpha$ is unknown and therefore often
assigned a prior distribution. In this case, the marginal posterior density of
the model parameters $\theta$ based on the current data $D$ and the historical
data $D_0$ is given by
\begin{align*}
    \pi(\theta \given D, D_0) = \int_0^1 \pi(\theta \given D, D_0, \alpha)  \pi(\alpha \given D, D_0)
     \text{d}\alpha,
\end{align*}
that is, the posterior of $\theta$ based on a fixed $\alpha$ averaged over the
marginal posterior of $\alpha$. The marginal posterior of $\alpha$ thus
determines how much pooling between the two data sets occurs, and a complete
pooling of both data sets happens when the posterior has all its mass at
$\alpha = 1$.

Standard Bayesian asymptotic theory establishes that, under certain regularity
conditions, posterior distributions become more concentrated with increasing
amounts of data \citep[section 5.3]{Bernardo2000}. Since the power prior is
based on the historical data, intuition would suggest that as the sample sizes
of the current and historical data sets increase, the marginal posterior of
$\alpha$ should become increasingly concentrated at $\alpha = 0$ if there is
conflict between the data sets, and increasingly concentrated at $\alpha = 1$ if
there is no conflict. Here, we show that only the former is true, but not the
latter, at least in the typical situation where $\alpha$ has a beta prior
distribution and the data have normal or binomial likelihood. In case of
conflict, there is instead a limiting posterior distribution that is hardly
different from the prior. Our results imply that complete discounting is
possible, but complete pooling is impossible.

This paper is structured as follows: We start by showing the claimed result for
parameter estimates under normality (Section~\ref{sec:normal}). We then show
that the result also approximately holds in the binomial model
(Section~\ref{sec:binomial}). The papers then ends with some concluding remarks
in Section~\ref{sec:discussion}. As a running example we consider data from two
randomized clinical trials comparing the efficacy of the drugs Fidaxomicin and
Vancomycin on \textit{Clostridium difficile}-associated diarrhoea in adults
(Table~\ref{tab:data}).

\begin{table}[!htb]
  \centering
  \caption{Historical and current data on the comparison between Fidaxcomicin and
  Vancomycin with respect to their effect on \textit{Clostridium
    difficile}-associated diarrhoea in adults. The number of participants and events
    were taken from the respective intention-to-treat analysis of the studies as in
    the meta-analysis of \citet{Nelson2017}.
  }
  \label{tab:data}

\begin{tabular}{cccc}
  \toprule
  Study & Risk (Fidaxomicin) & Risk (Vancomycin) & Risk ratio (95\% CI) \\
  \midrule
  \citet{Cornely2012} & $\dfrac{193}{270} = 71.5\%$
  & $\dfrac{163}{265} = 61.5\%$
  & $1.16$ (1.04, 1.31) \\
  & & & \\
  \citet{Louie2011} & $\dfrac{214}{302} = 70.9\%$
  & $\dfrac{198}{327} = 60.6\%$
  & $1.17$ (1.04, 1.31) \\
  \bottomrule
\end{tabular}
\end{table}

\section{Parameter estimates under normality}
\label{sec:normal}
Define the current data set by $D = \{\that, \sigma\}$ with $\that$ an estimate
of an unknown univariate parameter $\theta$ and $\sigma$ the (assumed to be
known) standard error of the estimate. Denote by $D_0 = \{\that_0, \sigma_0\}$
the respective quantities obtained from the historical data. The standard errors
are usually of the form $\sigma = \kappa/\sqrt{n}$ and
$\sigma_{0} = \kappa/\sqrt{n_{o}}$, where $n$ and $n_{0}$ are effective sample
sizes and $\kappa^{2}$ is the variance of one unit. The relative variance
$c = \sigma^{2}_{0}/\sigma^{2} = n/n_{0}$ can then be interpreted as a ratio of
sample sizes. For both estimates, assume a normal likelihood centered around the
parameter $\theta$ and with variance equal to the squared standard error. In the
default ``normalized'' version of the power prior \citep{Duan2005,
  Neuenschwander2009} the prior for the power parameter $\alpha$ is assigned
marginally to the posterior distribution of $\theta$ based on an initial prior
for $\theta$ and the likelihood of the historical data $D_0$ raised to the power
of $\alpha$. Here and henceforth we will use an initial uniform prior for the
parameter $\pi(\theta) \propto 1$ and a beta prior for the power parameter
$\alpha \sim \Be(p, q)$. This choice leads to the normalized power prior
\begin{align}
    \label{eq:nppnormal}
  \pi(\theta, \alpha \given D_0)
  = \frac{L(D_{0} \given \theta)^{\alpha} \pi(\alpha)}{\int_{-\infty}^{+\infty} L(D_{0} \given \theta^{\prime})^{\alpha}  \text{d}\theta^{\prime}}
    = \Nor(\theta\given \that_0, \sigma^2_0 /\alpha) \Be(\alpha \given p, q)
\end{align}
with $\Nor(\cdot \given m, v)$ the normal density function and
$\Be(\cdot \given p, q)$ the beta density function.
Combining~\eqref{eq:nppnormal} with the likelihood of the current data produces
a joint posterior for $\theta$ and $\alpha$, \ie{}
\begin{align*}
  \pi(\theta, \alpha \given D, D_0)
  &= \frac{L(D \given \theta)  \pi(\theta, \alpha \given D_{0})}{\int_{0}^{1} \int_{-\infty}^{\infty}
  L(D \given \theta^{\prime})  \pi(\theta^{\prime}, \alpha^{\prime} \given D_{0})
  \text{d}\theta^\prime \text{d}\alpha^\prime}
  = \frac{\Nor(\that \given \theta, \sigma^{2}) \Nor(\theta \given \that_0, \sigma^2_0 / \alpha) \Be(\alpha \given p, q)}{
    \int_0^1 \Nor(\that \given \that_0, \sigma^2 + \sigma^2_0 / \alpha^\prime) \Be(\alpha^\prime \given p, q)
    \text{d}\alpha^\prime},
\end{align*}
from which a marginal posterior for
$\alpha$ can be obtained by integrating out $\theta$, \ie{}
\begin{align}
    \label{eq:margpostnormal}
    \pi(\alpha \given D, D_0)
  = \int_{-\infty}^{+\infty} \pi(\theta, \alpha \given D, D_{0}) \text{d}\theta
  = \frac{\Nor(\that \given \that_0, \sigma^2 + \sigma^2_0 / \alpha) \Be(\alpha \given p, q)}{
    \int_0^1 \Nor(\that \given \that_0, \sigma^2 + \sigma^2_0 / \alpha^\prime) \Be(\alpha^\prime \given p, q)
    \text{d}\alpha^\prime}.
\end{align}

\begin{figure}[!htb]
\begin{knitrout}
\definecolor{shadecolor}{rgb}{0.969, 0.969, 0.969}\color{fgcolor}
\includegraphics[width=\maxwidth]{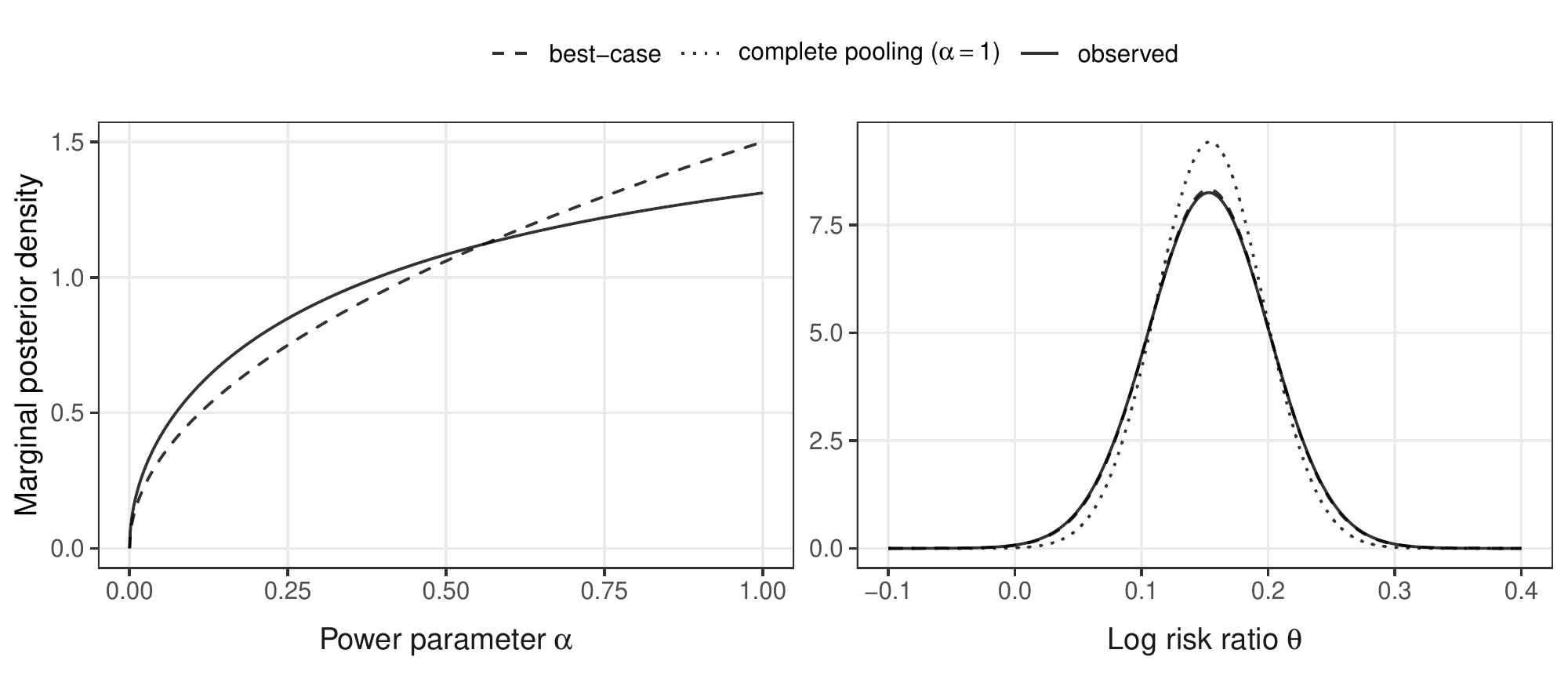} 
\end{knitrout}

\caption{Power prior modeling of current data
  $D = \{\that = 0.15, \sigma = 0.06\}$
  from \citet{Cornely2012} and historical data
  $D_{0} = \{\that_{0} = 0.16, \sigma_{0} = 0.06\}$
  from \citet{Louie2011}. A $\alpha \sim \Be(1, 1)$ prior is used for the power
  parameter. Marginal posterior densities are computed by numerical
  integration.}
  \label{fig:analysis}
\end{figure}

The black solid lines in Figure~\ref{fig:analysis} show the marginal posterior
distributions of the power parameter $\alpha$ (left) and the log risk ratio
$\theta$ (right) based on a uniform $\alpha \sim \Be(1, 1)$ prior for the power
parameter and the data from Table~\ref{tab:data}; the log risk ratio with
standard error from the current data $D = \{\that = 0.15, \sigma = 0.06\}$, and
the log risk ratio with standard error from the historical data
$D_{0} = \{\that_{0} = 0.16, \sigma_{0} = 0.06\}$. Both distributions are
computed via numerical integration.

The left plot of Figure~\ref{fig:analysis} shows that the observed marginal
posterior of $\alpha$ has hardly changed from the uniform prior, despite the
almost perfect correspondence of historical and current log risk ratio. The
dashed line shows the (soon to be discussed) marginal posterior for the
best-case scenario when both log risk ratios perfectly correspond and their
standard errors become arbitrarily small. As is visible, this best-case
posterior of $\alpha$ is not too far off the observed one, giving only slightly
more support to larger values of $\alpha$.

The right plot of Figure~\ref{fig:analysis} shows the corresponding marginal
posterior for $\theta$. As can be seen, the observed posterior (solid line) and
the best-case posterior (dashed line) are virtually indistinguishable,
suggesting that the data achieve as much pooling as the normalized power prior
model permits. In contrast, the dotted line indicates that the posterior based
on complete pooling of the data sets would be somewhat more peaked.

In general the integral in the denominator of~\eqref{eq:margpostnormal} has to
be computed by numerical integration, but there are certain important situations
when an analytical solution exists, and further insight can be gained. We will
discuss these cases in the following.

\subsection{Perfect compatibility of historical and current data}
The first situation occurs when the current data perfectly mirror the historical
data in the sense that both parameter estimates are equivalent
($\that = \that_0$). In this case, several terms cancel
in~\eqref{eq:margpostnormal} so that the integral can be represented in terms of
the hypergeometric function
${}_2F_1(a, b, c; z) = \{\int_0^1 t^{b-1}(1-t)^{c - b -1} (1 - tz)^{-a} \text{d}t\}/\mathrm{B}(b, c - b)$
with $\mathrm{B}(x, y)$ the beta function \citep[section
15.3.1]{Abramowitz1964}, \ie{}
\begin{align}
    \label{eq:margpostsimilar}
    \pi(\alpha \given D, D_0, \that = \that_0)
    % &= \frac{(1/c + 1/\alpha)^{-1/2} \times \Be(\alpha\given p, q)}{
    % \int_0^1 (1/c + 1/\alpha^\prime)^{-1/2} \times \Be(\alpha^\prime \given p, q)
    % \text{d}\alpha^\prime} \\
    &= \frac{(\alpha/c + 1)^{-1/2}  \Be(\alpha \given p + 1/2, q)}{
    {}_2F_1(1/2, p + 1/2, p + q + 1/2; -1/c)}.
\end{align}
where $c = \sigma^2_0/\sigma^2$ is the relative variance. The
distribution~\eqref{eq:margpostsimilar} is close to a $\Be(p + 1/2, q)$
distribution which is hardly different from the $\Be(p, q)$ prior distribution,
despite perfect compatibility of both data sets. Importantly, the marginal
posterior~\eqref{eq:margpostsimilar} does not depend on the actual value of the
standard errors $\sigma$ and $\sigma_0$ but only on the relative variance $c$.
This means that~\eqref{eq:margpostsimilar} holds for finite standard errors but
also in the idealized mathematical situation where both standard errors go
equally fast to zero (\ie{} infinite sample size), but with possibly different
starting values ($c\neq 1$).

Typically, the historical data are predetermined and only the standard error of
the current study can be changed. It is therefore interesting to study the
behavior of~\eqref{eq:margpostsimilar} for $c \to \infty$, \ie{} the current
standard error $\sigma$ goes to zero while the historical standard error
$\sigma_0$ remains fixed, reflecting an arbitrary increase of the current sample
size. In that case it is straightforward to see from the power series
representation of the hypergeometric function \citep[section
15.1.1]{Abramowitz1964} that
\begin{align*}
    \lim_{c\to\infty} {}_2F_1(1/2, p + 1/2, p + q + 1/2; -1/c)
    &= \lim_{c\to\infty} 1 + \mathcal{O}(1/c)= 1.
\end{align*}
Hence, the limiting posterior density is
\begin{align}
    \label{eq:limitpostnormal1}
    \lim_{c\to \infty} \pi(\alpha \given D, D_0, \that = \that_0)
    &= \Be(\alpha \given p + 1/2, q),
\end{align}
that is, again a beta density but with updated success parameter $p + 1/2$, so
just slightly different from the prior. The limiting $\Be(3/2, 1)$ distribution
for the uniform prior is depicted by the dashed line in the left plot of
Figure~\ref{fig:analysis}.

\subsection{Arbitrarily precise current data}
The second situation in which the marginal posterior~\eqref{eq:margpostnormal}
is available in closed form is the limiting case when the current standard error
$\sigma$ goes to zero while the historical standard error $\sigma_0$ remains
fixed (but in contrast to the previous situation the parameter estimates $\that$
and $\that_0$ can take different values). In this case, the integral
in~\eqref{eq:margpostnormal} can be represented by the confluent hypergeometric
function
\mbox{$M(a, b, z) = \{\int_0^1 \exp(zt) t^{a-1}(1-t)^{b-a-1} \text{d}t\}/\mathrm{B}(b - a, a)$}
\citep[section 13.2.1]{Abramowitz1964} so that the marginal posterior is given
by
\begin{align}
     \label{eq:limitpostnormal2}
     \lim_{\sigma \downarrow 0} \pi(\alpha \given D, D_0)
     &=
     \frac{\exp\left\{-\alpha  (\that - \that_0)^2/(2\sigma^2_0)\right\}}{
     M\{p + 1/2, p + q + 1/2, -(\that - \that_0)^2/(2\sigma^2_0)\}}
     \Be(\alpha \given p + 1/2, q).
\end{align}
As expected, the distribution~\eqref{eq:limitpostnormal2} reduces
to~\eqref{eq:limitpostnormal1} when the parameter estimates are equal
($\that = \that_0$) since then the left fraction becomes one, which can be shown
using the power series representation of the confluent hypergeometric function
\citep[section 13.1.2]{Abramowitz1964}.

\begin{figure}[!htb]
\begin{knitrout}
\definecolor{shadecolor}{rgb}{0.969, 0.969, 0.969}\color{fgcolor}
\includegraphics[width=\maxwidth]{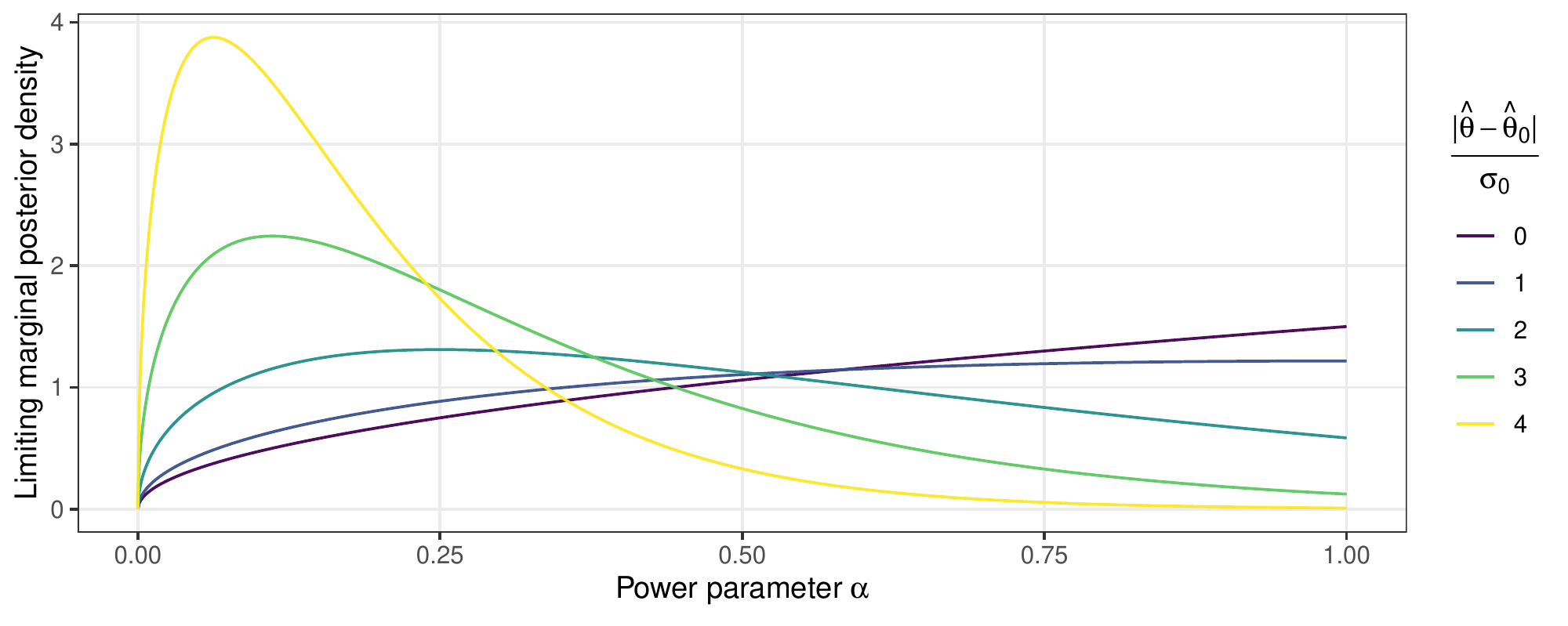} 
\end{knitrout}

\caption{Limiting marginal posterior distribution of power parameter $\alpha$
  based on $\alpha \sim \Be(1, 1)$ prior and when the current standard error
  goes to zero ($\sigma \downarrow 0$), for different values of the parameter
  difference standardized by the historical standard error
  $|\that- \that_0|/\sigma_0$.}
  \label{fig:limitingpost}
\end{figure}

Figure~\ref{fig:limitingpost} shows the distribution~\eqref{eq:limitpostnormal2}
for different values of the parameter estimate difference standardized by the
historical standard error ($|\that-\that_0|/\sigma_0$). One can see that when
the parameter estimates become more different (larger $|\that_0 -\that|$) the
limiting distribution~\eqref{eq:limitpostnormal2} will be increasingly shifted
towards smaller values of $\alpha$ indicating more incompatibility among the
data sets and reducing the information borrowing from the historical data. This
shift amplifies with decreasing historical standard errors $\sigma_0$, meaning
that the posterior can become arbitrarily peaked by increasing the sample size
of the historical study. In contrast, when the parameter estimates are the same
($\that = \that_0$) the historical standard error $\sigma_0$ does not influence
the posterior.

The marginal posterior for $\alpha$ can thus at best become a $\Be(p + 1/2, q)$
if the prior is a $\Be(p , q)$, implying that a complete pooling of historical
and current data can never be achieved. On the other hand, a complete
discounting is possible since conflict between the current and historical data
can make the marginal posterior arbitrarily peaked at $\alpha = 0$. However,
considering that the distribution~\eqref{eq:limitpostnormal2} is based on an
extremely informative current data set which lead to an estimate of the unknown
parameter $\theta$ without any measurement error, the rate at which the
posterior becomes more concentrated seems also dissatisfactory. For instance,
for a standardized parameter difference $|\that- \that_0|/\sigma_0 = 3$, the
posterior is only slightly peaked at around $\alpha = 0.1$.

\section{Binomial model}
\label{sec:binomial}
So far we assumed a normal likelihood and a univariate parameter,
Appendix~\ref{app:normallinear} shows that the previous result can be
generalized to normal linear models with multivariate parameter $\theta$. In
this section, we will show that the previous result also holds approximately in
the binomial model, which is an important model class in medical applications of
power priors.

Let $D = \{x, n\}$ and $D_0 = \{x_0, n_0\}$ denote the number of successes and
total trials from current data and historical data set, respectively. Assume a
binomial likelihood
% $X \given \theta \sim \Bin(n, \theta)$
with success probability $\theta$ for each of them, and let $\hat{\theta} = x/n$
and $\hat{\theta}_0 = x_0/n_0$ denote the respective maximum likelihood
estimates. Assigning an initial beta prior $\theta \sim \Be(0, 0)$ for the
success probability and a beta prior $\alpha \sim \Be(p, q)$ for the power
parameter leads to the normalized power prior
\begin{align}
  \label{eq:priorbin}
  \pi(\theta, \alpha \given D_0)
  &= \Be\{\theta \given \alpha x_0, \alpha (n_0 - x_0)\}
  \Be(\alpha \given p, q).
\end{align}
Combining the prior~\eqref{eq:priorbin} with the likelihood of the current data
leads to a joint posterior distribution for $\theta$ and $\alpha$, from which a
marginal posterior for $\alpha$ can be obtained by integrating out $\theta$,
\ie{}
\begin{align}
  \label{eq:margpostalpha}
  \pi(\alpha \given D, D_0)
  = \frac{\BetaBin\{x \given n, \alpha  x_0, \alpha (n_0 - x_0)\}
  \Be(\alpha \given p, q)}{\int_0^1
  \BetaBin\{x \given n, \alpha^\prime   x_0, \alpha^\prime (n_0 - x_0)\}
  \Be(\alpha^\prime \given p, q) \text{d}\alpha^\prime}
\end{align}
with % $\Bin(\cdot \given n, \theta)$ the binomial probability mass function and
$\BetaBin(\cdot \given n, p, q)$ the beta-binomial probability mass function.

As for the normal model, the integral in the denominator
of~\eqref{eq:margpostalpha} is generally not available in closed form. Yet
again, it is possible to obtain a closed form expression when the probability
estimates from both studies are equivalent ($\hat{\theta} = \hat{\theta}_0$).
Application of Stirling's approximation
$\mathrm{B}(x, y) \approx \sqrt{2\pi} x^{x-1/2}y^{y-1/2}(x + y)^{-(x+y-1/2)}$ to both
beta functions in the probability mass function of the beta-binomial leads then
to
\begin{align}
     \label{eq:betaratio}
     \BetaBin\{x \given n, \alpha  x_0, \alpha (n_0 - x_0)\}
  &\approx \binom{n}{x}  \that^x (1 - \that)^{n - x}
  \sqrt{\frac{\alpha n_0}{n + \alpha n_0}}.
\end{align}
Using the approximation~\eqref{eq:betaratio} in numerator and denominator
of~\eqref{eq:margpostalpha} produces the same marginal
posterior~\eqref{eq:margpostsimilar} as for the normal model but with
$c = n/n_0$, representing the relative sample size of the data sets. The
limiting marginal posterior distribution of $\alpha$ is thus (approximately) the
same for normal and binomial model.

\begin{figure}[!htb]
\begin{knitrout}
\definecolor{shadecolor}{rgb}{0.969, 0.969, 0.969}\color{fgcolor}
\includegraphics[width=\maxwidth]{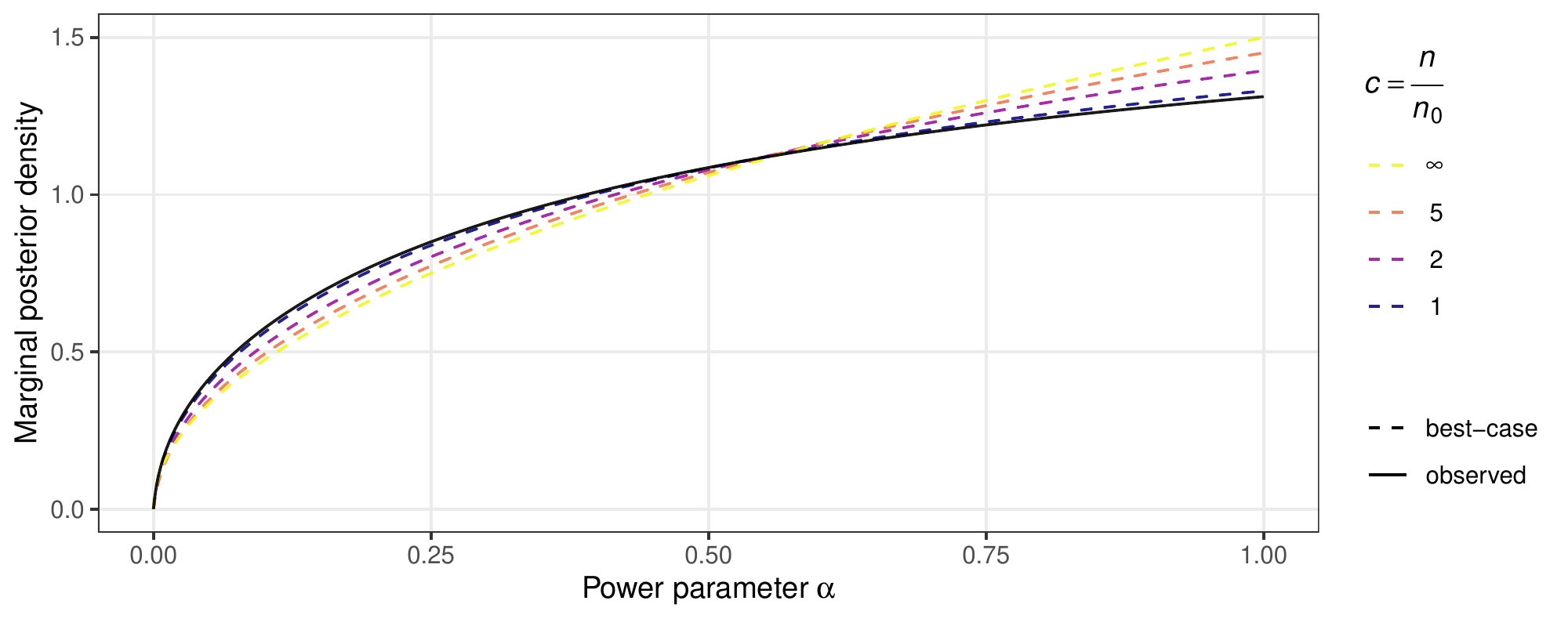} 
\end{knitrout}
\caption{Marginal posterior of power parameter $\alpha$ based on
  $\alpha \sim \Be(1, 1)$ prior and historical data
  $D_{0} = \{x_{0} = 214, n_{0} = 302\}$ from \citet{Louie2011}.
  The black solid line shows the marginal posterior for the actually observed
  current data $D = \{x = 193, n = 270\}$ from \citet{Cornely2012},
  whereas the dashed lines show the best-case marginal posteriors for
  hypothetical current data $D = \{x = c \times x_{0}, n = c \times n_{0}\}$
  which perfectly mirror the original data but with relative sample sizes
  $c = n/n_{0}$.}
\label{fig:binomial}
\end{figure}

Figure~\ref{fig:binomial} shows the marginal posterior~\eqref{eq:margpostalpha}
based on the data from Table~\ref{fig:analysis} using the risk in the
Fidaxomicin group as the parameter of interest. The black solid line depicts the
posterior based on the observed data from \citet{Cornely2012}, whereas the
dashed lines depict posteriors based on hypothetical data which perfectly mirror
the historical data (\ie{} with identical probability estimates
$\that = \that_0$) from \citet{Louie2011} but with different relative sample
sizes $c = n/n_0$. Numerical integration is used for computing the posterior in
all finite sample size cases, for $c\to \infty$ the limiting
distribution~\eqref{eq:limitpostnormal1} is shown. It can be seen that the
limiting posterior density based on Stirling's approximation (yellow line) is
close to the exact posterior for finite relative sample sizes $c$, suggesting
that the approximation is reasonably accurate. Moreover, the observed marginal
posterior (which has $c = 0.89$) is very close to the best-case posterior for
$c = 1$, indicating that the data achieve almost as much pooling as the power
prior permits in that case.

\section{Discussion}
\label{sec:discussion}
We showed that normalized power priors in normal and binomial models combined
with beta priors assigned to the power parameter $\alpha$ have undesirable and
counterintuitive properties. Specifically, in the best-case scenario when the
current data perfectly mirror the historical data and the sample sizes from both
data sets become arbitrarily large, the marginal posterior of $\alpha$ does not
converge to a point mass at $\alpha = 1$ but approaches a
$\alpha \sim \Be(p + 1/2, q)$ distribution, hardly differing from the prior
$\alpha \sim \Be(p, q)$. The result implies that a complete pooling of
historical and current data can never be achieved. Our case study illustrates
that the property is not only a mathematical curiosity but can occur in
statistical analysis of medical data.

We still believe that normalized power priors are useful since they permit
arbitrarily large data-driven discounting of historical data. However, data
analyst should be aware that this does not work in the other direction as the
amount of possible pooling is predetermined by the prior. Data analysts have
different options to alleviate this limitation. For instance, they can use the
power prior based on a fixed $\alpha$ and use either a ``guide value''
\citep{Ibrahim2015} or elicit a reasonable value from external knowledge about
the similarity of the data sets. Another option is to specify informative priors
which give most of their mass to larger values of $\alpha$, thereby shifting the
best-case marginal posterior to larger values as well. Finally, a pragmatic
alternative is to specify $\alpha$ via an empirical Bayes approach as proposed
by \citet{Gravestock2017}, which permits complete pooling of both data sets.

We only studied the limiting marginal posterior of $\alpha$ in the normal and
binomial models combined with beta priors on $\alpha$, yet we conjecture that
the issue is more fundamental and also present in other types of models.
However, this will likely be more difficult to establish as marginal posteriors
are typically not available in closed form for more complex models.

For the normal model, there is an exact correspondence between power parameter
models with fixed $\alpha$ and hierarchical (random-effects meta-analysis)
models with fixed heterogeneity variance \citep{Chen2006}. This connection may
provide an intuition for why the counterintutive result occurs: Precisely
estimating a heterogeneity variance from two observations alone (the historical
and current data sets) seems like an impossible task as the ``unit of
information'' is the number of data sets and not the number of samples within a
data set. We report elsewhere on the precise connection between power parameter
and hierarchical models when power parameter and the heterogeneity variance are
random \citep{Pawel2022ppfrs}.

\section*{Software and data}
The summary data used in this study were extracted from Figure~5.1 in
\citet{Nelson2017}. All analyses were conducted in the R programming language
version 4.3.0 \citep{R}. The
packages \texttt{ggplot2} \citep{Wickham2016} and \texttt{hypergeo}
\citep{Hankin2016} were used used for graphics and (confluent) hypergeometric
function implementation, respectively. The code and data to reproduce our
analyses is openly available at \url{https://github.com/SamCH93/ppPooling}. A
snapshot of the GitHub repository at the time of writing is available at
\url{https://doi.org/10.5281/zenodo.6626963}.

\section*{Acknowledgments}
This work was supported in part by an NWO Vici grant (016.Vici.170.083) to EJW,
an Advanced ERC grant (743086 UNIFY) to EJW, and a Swiss National Science
Foundation mobility grant (189295) to LH and SP.

\begin{appendices}
\section{Normal linear model}
\label{app:normallinear}
Define the current data by $D = \{\yb, \Xb\}$ with $\yb$ an $n \times 1$ vector
of observations and $\Xb$ an $n \times k$ matrix of known covariates. Assume a
linear model $\yb = \Xb \thetab + \varepsilonb$ with $\thetab$ a $k \times 1$
vector of parameters and $\varepsilonb$ an $n \times 1$ vector of random errors
whose components are independently normally distributed around zero with (known)
variance $\sigma^2$ variance. Denote by $D_0 = \{\yb_0, \Xb_0\}$ the respective
quantities from a historical data set. Finally, let
$\hat{\thetab} = (\Xbt \Xb)^{-1} \Xbt \yb$ and
$\hat{\thetab}_0 = (\Xbt_0 \Xb_0)^{-1} \Xbt_0 \yb_0$ denote the respective
maximum likelihood estimates of $\thetab$. Based on a uniform initial prior
$\pi(\thetab) \propto 1$ for the parameter vector and a beta prior
$\alpha \sim \Be(p, q)$ for the power parameter, we obtain the normalized power
prior
\begin{align}
    \label{eq:npplinearmodel}
    \pi(\thetab, \alpha \given D_0)
    &= \Nor_\text{k}\left(\thetab \given \hat{\thetab}_0, \sigma^2  \alpha^{-1}
    (\Xbt_0 \Xb_0)^{-1}\right) \Be(\alpha \given p, q)
\end{align}
with $\Nor_\text{k}(\cdot\given \boldsymbol{\mu}, \boldsymbol{\Sigma})$ the
density function of a $k$-variate normal. Updating the
prior~\eqref{eq:npplinearmodel} with the likelihood of the current data leads to
a joint posterior for $\thetab$ and $\alpha$, from which a marginal posterior
for $\alpha$ can be obtained by integrating out $\thetab$, \ie{}
\begin{align}
    \label{eq:postmarg}
    \pi(\alpha \given D, D_0) =
    \frac{\Nor_\text{k}\left(\hat{\thetab} \given \hat{\thetab}_0, \sigma^2
    (\Xbt \Xb)^{-1} + \sigma^2  \alpha^{-1} (\Xbt_0 \Xb_0)^{-1}\right)
    \Be(\alpha \given p, q)}{\int_0^1
    \Nor_\text{k}\left(\hat{\thetab} \given \hat{\thetab}_0, \sigma^2
    (\Xbt \Xb)^{-1} + \sigma^2  {\alpha^\prime}^{-1}   (\Xbt_0 \Xb_0)^{-1}\right)
    \Be(\alpha^\prime \given p, q)  \text{d}\alpha^\prime}.
\end{align}
When the current and historical data are perfectly compatible (characterized by
equivalent maximum likelihood estimates $\hat{\thetab} = \hat{\thetab}_0$), the
integral in~\eqref{eq:postmarg} can again be represented in terms of the
hypergeometric function, and the marginal posterior of $\alpha$ is given
by~\eqref{eq:margpostsimilar} but with $c = |\Xbt \Xb|/|\Xbt_0 \Xb_0|$ the ratio
of the determinants of the precision matrices $(\Xbt \Xb)/\sigma^2$ and
$(\Xbt_0 \Xb_0)/\sigma^2$ of the maximum likelihood estimates $\hat{\thetab}$
and $\hat{\thetab}_0$.

\end{appendices}

\bibliographystyle{apalikedoiurl}
\bibliography{bibliography.bib}

\end{document}